# Broadband Absorbers and Selective Emitters based on Plasmonic Brewster Metasurfaces

Christos Argyropoulos[1], Khai Q. Le[1], Nadia Mattiucci[2], Giuseppe D'Aguanno[2], Andrea Alù[1,*]

[1]Dept. of Electrical & Computer Engineering, The University of Texas at Austin

[2]AEgis Tech., Nanogenesis Division, Huntsville, AL 35806, USA

*alu@mail.utexas.edu

*We discuss the possibility of realizing utlrabroadband omnidirectional absorbers and angularly selective coherent thermal emitters based on properly patterned plasmonic metastructures. Instead of relying on resonant concentration effects that inherently limit the bandwidth, we base our design on the combination of two inherently nonresonant effects: plasmonic Brewster funneling and adiabatic plasmonic focusing. With this approach, we demonstrate compact, broadband absorption and emission spanning terahertz, infrared and optical frequencies, ideal for various energy and defense applications.*

Infrared (IR) and optical emitters are currently based on semiconductor devices, such as light-emitting diodes (LEDs) and quantum cascade lasers [1]-[2]. In general, their operation is inherently narrowband, fabrication is quite challenging and their overall efficiency and emissivity are limited. In the last few years, thermal sources were suggested as an alternative way to overcome many of these limitations. An ideal thermal emitter follows Planck's law of blackbody radiation [3], but conventional thermal sources in general exhibit smaller and more narrowband emission than a blackbody, depending on material and design constraints. In addition, thermal sources emit incoherently over a relatively broad angular range, whereas in many applications spatial coherence and narrow emission angles are very desirable. Substantial research efforts have been recently dedicated to overcome these limitations and tame blackbody, angularly coherent emission features. One of the most promising venues is based on plasmonic and metamaterial effects [4]: resonant plasmonic gratings and metamaterials can significantly boost emission at the desired wavelength, up to the level of an ideal black-body, but usually this comes at the price of stringent bandwidth constraints [5]-[6]. Resonant nanoantenna emitters [7]-[8], periodic grooves [9], metallic gratings [10], microstrip patches [11] and wire medium [12] have also been successfully employed for similar purposes, with analogous limitations based on their resonant mechanisms. Since the emission properties of a device in thermodynamic equilibrium are directly related to its absorption features [13], many of these solutions have also been explored to realize narrowband absorbers and filters [14]-[18]. Different from emitters, in the case of absorbers a large acceptance angle may be desirable, where instead resonant solutions based on plasmon polaritons often lead to a sensitive response to the incident angle [19]-[20].

Realizing an ideally broadband blackbody response in a compact device with the added capability of tailoring its angular emission would open groundbreaking venues in sources, emitters and absorbers technology for a variety of energy and defense applications. In this regard, a few recent efforts have been devoted to realizing broadband operation based on tapered plasmonic geometries or combined multiple resonances [21]-[25]. These solutions still show bandwidth limitations, as they fundamentally rely on resonant processes. In this Letter, on the contrary, we rely on strictly nonresonant phenomena to achieve ultrabroadband emission and absorption with controllable angular selectivity, spanning with a single device THz, IR and visible frequencies. Our concept is based on the combination of two nonresonant effects: plasmonic Brewster light funneling at a single interface [26]-[27] and adiabatic plasmonic focusing [28].

Consider a one-dimensional (1D) grating with period $d$, formed by an array of slits carved in gold and infinitely extended along $y$ with unit cell shown in Fig. 1(a). The slits have width $w$ and length $l$, terminated by a taper with length $l_{tap}$ designed to adiabatically dissipate the energy transmitted through the slits [28]-[29]. The taper is then terminated by a gold back plate much thicker than the skin depth, ensuring that an external impinging wave can only be reflected or absorbed by the structure. We assume a relative Drude permittivity $\varepsilon_{Au} = \varepsilon_\infty - f_p^2 / [f(f+i\gamma)]$ for gold with parameters $f_p = 2069$ THz, $\gamma = 17.65$ THz, $\varepsilon_\infty = 1.53$ [30], under an $e^{-i\omega t}$ time dependence. In order to model the varying taper width, which becomes rapidly comparable to the electron mean free path in the metal [31], and of the temperature dependence of the gold resistivity [16], we assume an increased collision frequency in the taper portion $\gamma_{tap} = 176.5$ THz. This assumption is not necessary for the proposed concept, as discussed in [32], but is considered here to make our model more realistic.

As originally shown in [26], the grating interface may be tailored to be perfectly impedance matched to a transverse-magnetic (TM) impinging plane wave, provided that the incident angle satisfies the condition:

$$\cos\theta_B = \frac{\beta_s w}{\varepsilon_s k_0 d}, \qquad (1)$$

where $\varepsilon_s$ is the material permittivity filling the slits, $k_0 = \omega/c$ is the free-space wave number, $c$ is the velocity of light in free-space and $\beta_s$ is the guided wave number in a parallel-plate plasmonic slit with thickness $d$. At this angle, similar to the Brewster condition for a homogeneous interface, zero reflection and total transmission through the interface are expected [26]. Interestingly, this phenomenon weakly depends on frequency, as long as the plasmonic mode in the slit is weakly dispersive. This simple analytical model is a very accurate description of the anomalous funneling mechanism through the slits as long as the impinging wavelength is longer than $d$ [26], ensuring that the impinging energy can funnel into the slits from DC to very high frequencies, up to the wavelength $\lambda_0 \simeq d$. Independent confirmation of these findings was reported in [33]-[34].

This funneling phenomenon is purely based on impedance matching, without resorting on any resonance, and therefore the transmitted wave may be interestingly absorbed inside the slits without affecting at all the reflection coefficient or the bandwidth of operation. This functionality is very different from any other tunneling mechanism through narrow slits relying on resonant mechanisms, which would be severely affected by absorption. Absorption is achieved in our design by introducing a proper taper behind the Brewster interface, which can adiabatically absorb the transmitted plasmonic mode without reflections. The tapering angle and the corresponding length $l_{tap}$ determine, following [28]-[29],[35]-[36], the largest wavelength over which the transmitted energy gets fully absorbed in the metallic walls by the time it reaches the taper termination. Since the efficiency of adiabatic absorption depends on the taper length compared to the excitation wavelength, in fact, a given choice of tapering length fixes the limit on the minimum frequency of operation to achieve perfect absorption.

Figure 2 shows the absorption coefficient $A = 1 - R$, where $R$ is the calculated reflectance computed with the finite-integration method [37], for a grating as in Fig. 1(a) with $d = 96\,nm$, $w = 24\,nm$, $l = 200\,nm$, $\varepsilon_S = 1$ and $l_{tap} = 980\,nm$, consistent with the grating design proposed in [26]-[27] to support Brewster funneling at $\theta_B = 70^o$, as predicted by Eq. (1). As expected, around the Brewster angle total absorption may be achieved over a very broad range of wavelengths, effectively spanning from $\lambda_{min} \sim 200\,nm$ to $\lambda_{max} \sim 10\,\mu m$. Consistent with the previous discussion, this range may be further broadened, as the upper cut-off (shorter wavelength) is determined by the transverse period, whereas the lower limit is fixed by the taper length. The angle of total absorption is rather flat until $5-6\,\mu m$, and the angular range of absorption can be controlled by the ratio $d/w$ [27]. In this design, large absorption $A > 70\%$ is achieved for all incident angles in the frequency range of interest, even at normal incidence, except for angles very close to grazing incidence beyond the Brewster angle.

The proposed structure forms an omnidirectional, ultrabroadband absorber based on inherently nonresonant mechanisms. Compared to recent designs based on resonant effects [24],[38], we achieve broader absorption bandwidths and larger acceptance angles. Interestingly, by simply tailoring the design parameters as discussed above, we can further broaden both frequency and angular bandwidth as desired. In [32], we consider more lossy metals, such as platinum, which provide even better absorption features for similar total thickness. The proposed broadband omnidirectional absorber may mimic the absorption performance of an ideal blackbody with exciting applications in energy harvesting, absorbers and bolometers.

Seen these interesting absorption features, the same concept may be applied to realize black-body like thermal emission. The overall emissivity can be easily computed by multiplying its absorption coefficient by the Planck's law distribution of blackbody radiation [3]:

$$B_\lambda(T) = \frac{2hc^2}{\lambda^5} \frac{1}{e^{hc/\lambda k_B T} - 1},\qquad(2)$$

where $h$ is the Planck's constant, $k_B$ the Boltzmann constant and $T$ the absolute temperature. In this scenario, angular selectivity may be desirable, in order to channel the emitted energy towards specific directions in free-space and not waste it into other directions. Again, by simply tailoring the ratio $w/d$, we may be able to realize ultrabroadband emission with much sharper angular directivity.

We consider an emitter with dimension $d = 1.44\,\mu m$, $w = 90\,nm$, $l = 2\,\mu m$, $l_{tap} = 9.1\,\mu m$, $\varepsilon_S = 1$. First, we assume an operating temperature $T = 700\,K$, for which $B_\lambda$ is centered in the IR range, spanning an emission bandwidth shown in the bottom part of Fig. 3(a). The panel shows the thermal emission as a function of frequency and angle, normalized to the maximum value of blackbody radiation at this temperature. We essentially achieve an emission bandwidth equal to the one of an ideal black-body, with much large angular selectivity confined in a narrow beamwidth around the Brewster angle $\theta_B = 84^o$, with large, and tailorable by design, spatial coherence, ideal for a variety of applications, including thermo-photovoltaic and energy. It is very remarkable that the emission angle is very flat over the whole emission range, of great practical interest. The emitter does not 'waste' energy in the entire angular spectrum, as an omnidirectional ideal blackbody would, but instead routes it only towards a specific direction of interest. Its large spatial coherence is another important feature for IR sources [5].

It is important to remark that the operational bandwidth of the grating is broader than the blackbody spectrum at the temperature of interest, which fundamentally determines the emission bandwidth in Fig. 3a. This implies that the same structure can emit different IR frequencies by simply changing the temperature of operation! For example, in Fig. 3(b) we demonstrate thermal emission for the same grating at room temperature $T = 300\,K$. The structure emits with the same selective properties, at the same angle, but a different, redshifted IR spectrum, due to the lower temperature. By increasing the operating temperature to, e.g., $T = 3000\,K$ it may be possible to extend these concepts to optical emission. In this case, gold would melt at these high temperatures and other metals, such as tungsten [16] may be alternatively considered, as we discuss in [39].

So far, the proposed geometries are 1D configurations, operating only for TM excitation in the plane of the grating. We have recently extended the Brewster funneling concepts to 2D [40], showing that a mesh of orthogonal slits may provide light funneling independent of the plane of polarization. We show the corresponding design for 2D absorbers and emitters in Fig. 1(b). The structure is formed by crossed slits, tapered in 2D to allow adiabatic focusing and absorption (and reciprocally emission) on all planes of TM polarization. In order to test the performance of this device in both absorption and emission, we analyze their functionality in the worst-case scenario of an azimuthal angle $\phi = 45^o$ between the two orthogonal sets of slits. The structure is expected to perform equal or better than its functionality at this angle.

Figure 4(a) compares the performance (Fig. 2) of the 1D absorber in Fig. 1(a) at normal incidence and at the Brewster angle $\theta_B = 70^o$ with the equivalent 2D case monitored on the $\phi = 45^o$ plane (which is the worst-case scenario), obtained by simply introducing another set of orthogonal slits with same period and width, as shown in Fig. 1(b). We notice that the 2D device has remarkable very similar performance with the 1D case, extending its functionality to all polarization planes. As expected, both designs show very large, broadband absorption, especially large at the Brewster angle (red lines), but consistently large for any angle, even at normal incidence (black lines). The design of Fig. 1(b) effectively provides an omnidirectional ultrabroadband absorber extending from far-IR to the optical range, and on all planes of polarization.

Figure 4(b) shows a similar comparison for a 2D broadband selective emitter with the same slit and grating dimensions as in Fig. 3 and for $T = 700\,K$. Again, very good agreement is obtained between the two structures, with a good contrast between emission at the Brewster angle compared to normal emission. This ensures that the 2D grating can operate as a broadband, angularly selective coherent thermal emitter, producing in this configuration a 2D conical directive radiation focused at the Brewster angle. Very exciting applications are envisioned based on the proposed devices, such as directive thermo-photovoltaic radiation.

To conclude, we have proposed here a novel concept to realize ultrabroadband omnidirectional absorbers and emitters based on 1D and 2D plasmonic gratings. Plasmonic Brewster funneling of energy and adiabatic focusing and absorption in tapered plasmonic slits have been combined to achieve large nonresonant absorption features over a controllable angular range, offering the possibility to realize omnidirectional absorbers, but also spatially coherent and angularly selective emitters. Design parameters can easily control the performance of our device: the grating period controls the upper frequency cut-off and the taper length, or equivalently the depth of our structure, controls the lower frequency of operation, whereas the angular spectrum is determined by the ratio $w/d$. For emission, the spectrum is fundamentally controlled by the temperature, as in the case of an ideal blackbody. We notice that the limitation on the minimum frequency of operation on the device thickness is consistent with the fundamental bounds on thickness to bandwidth ratio [41] for radar absorbers.

The proposed structures may increase the efficiency of optical and IR energy harvesting devices and lead to novel bolometer designs. Our recent experimental setup [42] for Brewster microwave funneling, may be used to realize similar effects also for high-gigahertz emission and absorption. Finally, the proposed thermal emitters with small angular beamwidth may lead to novel directional IR thermal sources with broadband coherent emission, as well as novel thermo-photovoltaic screens for efficient heat conversion and radiation. These thin covers may be realized with nanoskiving [43] or related nanofabrication techniques. More efficient solar cells may be built based on the proposed concepts, generally leading to new robust photonic devices.

This work has been supported by the ARO STTR project "Dynamically Tunable Metamaterials", the AFOSR YIP award No. FA9550-11-1-0009, the ONR MURI grant No. N00014-10-1-0942 and the DARPA SBIR project "Nonlinear Plasmonic Devices".

**Figures**

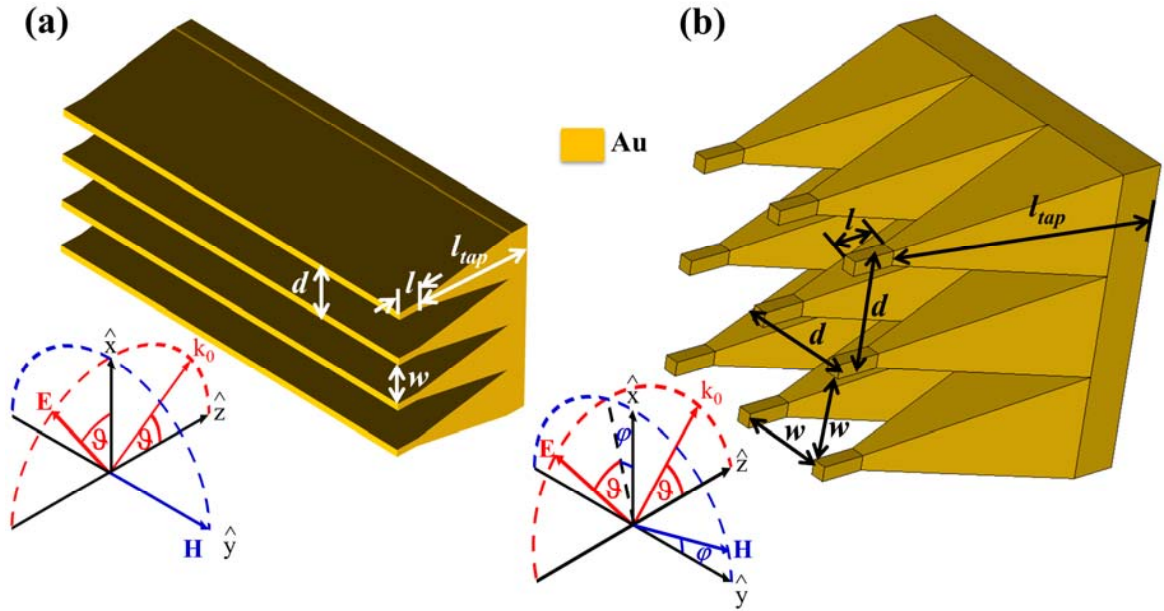

Figure 1 – a) Geometry of a 1D periodic grating combined with a plasmonic taper. b) Geometry of a 2D grating obtained combining two crossed plasmonic tapers. Both devices are illuminated at oblique incidence by a TM polarized wave. Notice that these model sketches are not in scale.

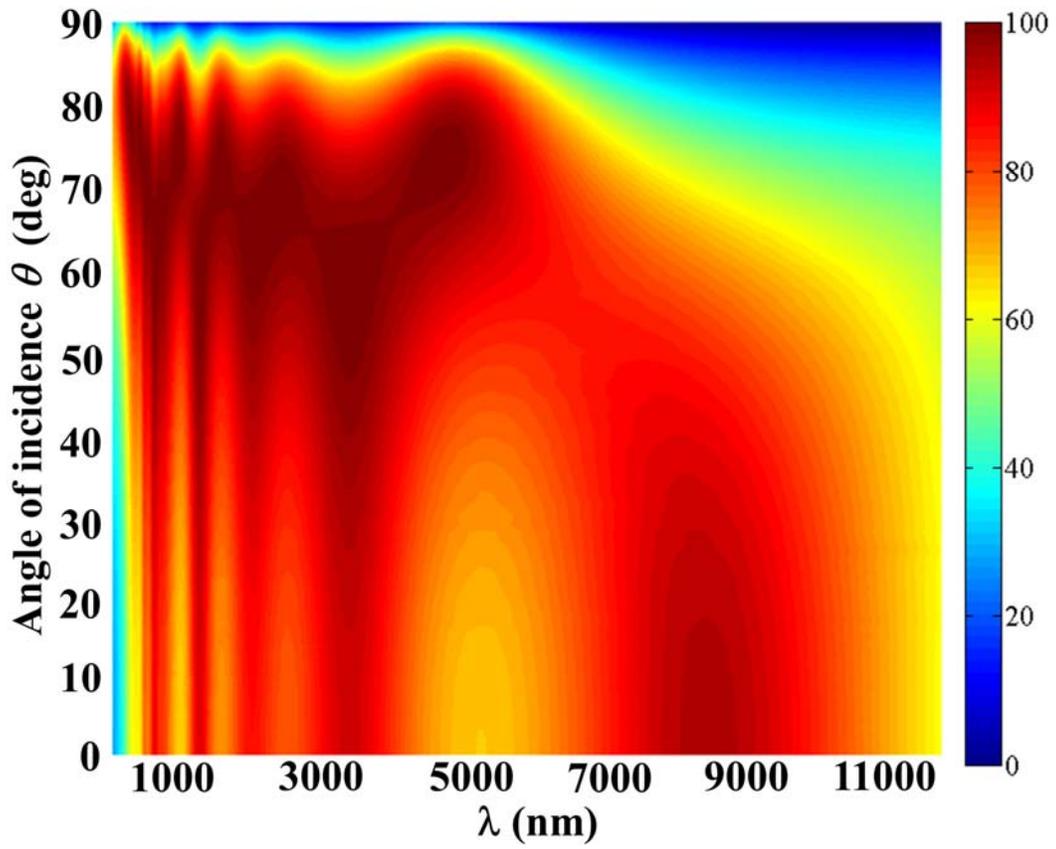

Figure 2 – Angular absorption spectra for the structure in Fig. 1(a) with period $d = 96\,nm$ and slit width $w = 24\,nm$ with $\varepsilon_s = 1$. The grating and taper lengths are $l = 200\,nm$ and $l_{tap} = 980\,nm$, respectively. Broadband omnidirectional absorption is obtained at optical and IR frequencies.

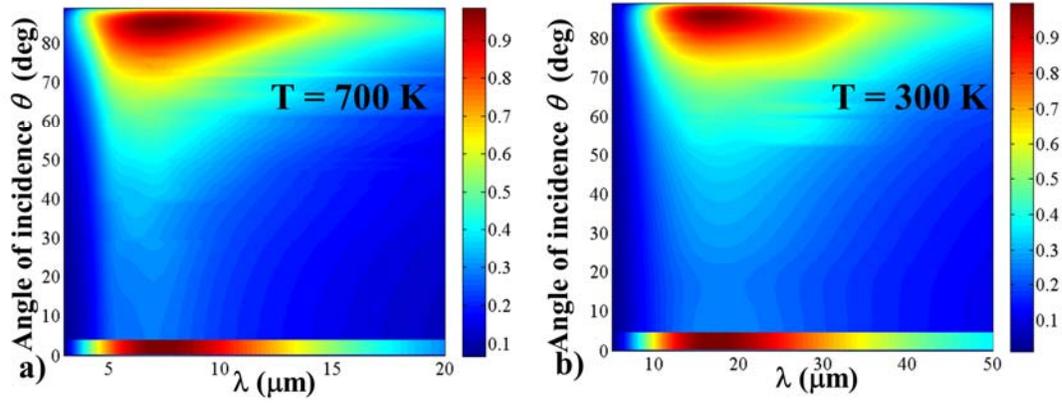

Figure 3 – Thermal emission of a directional plasmonic emitter with $d = 1.44\mu m$, $w = 90nm$, $l = 2\mu m$, $l_{tap} = 9.1\mu m$, $\varepsilon_s = 1$ at a) T = 700 K and b) T = 300 K. The insets in the bottom of the two panels show the emission spectrum of an ideal blackbody at the same temperature. Both plots are normalized to the maximum blackbody radiation.

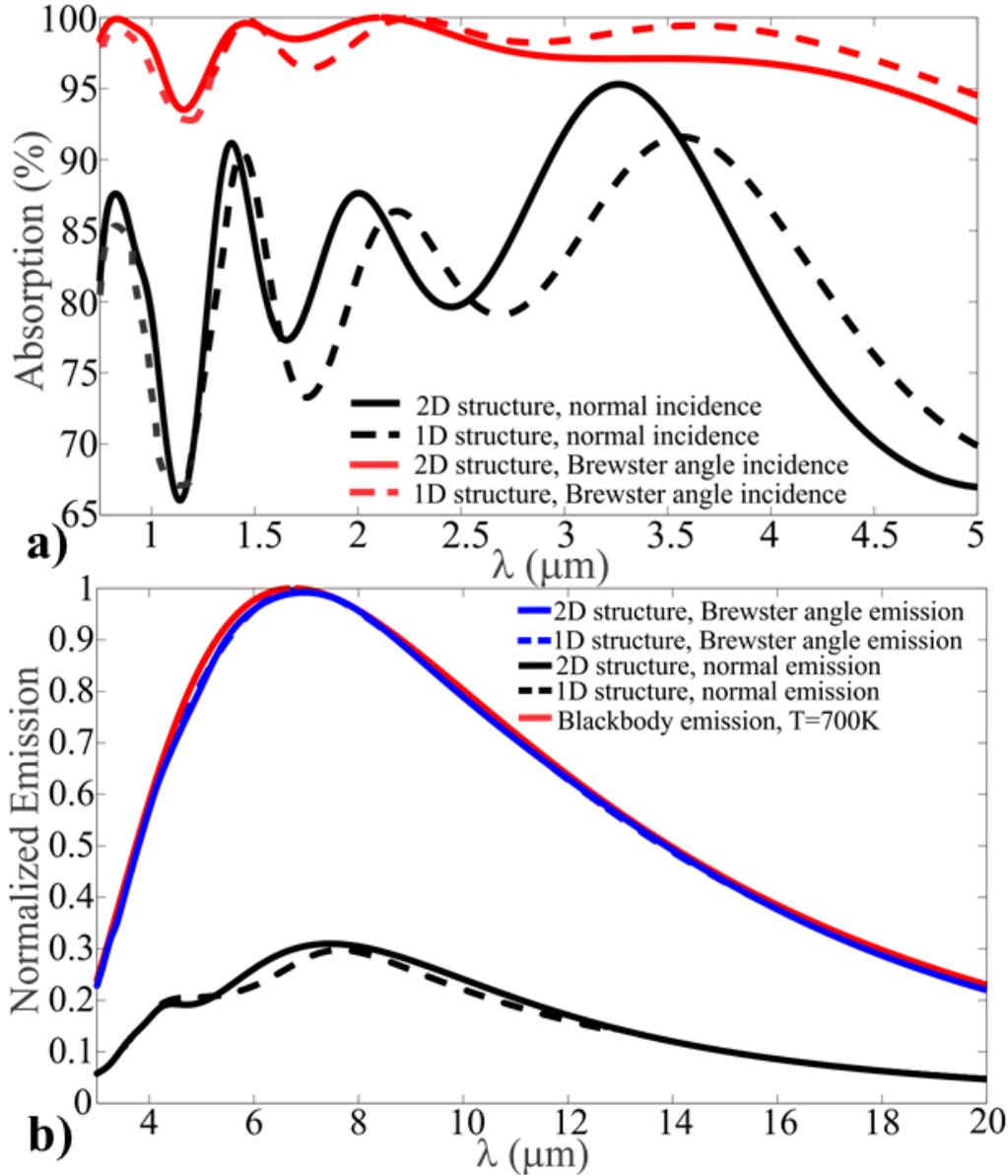

Figure 4 – a) Comparison of absorption for the 1D [Fig. 1(a)] and 2D [Fig. 1(b)] structures with same dimensions as in Fig. 2, both at the Brewster angle and at normal incidence. b) Similar comparison for emission at $T = 700\,K$ for the design in Fig. 3. In the 2D geometries, the azimuthal angle is fixed to $\phi = 45^o$, which represents the worst-case scenario.

# Supplementary material for the paper 'Broadband Omnidirectional Absorbers and Selective Thermal Emitters based on Plasmonic Brewster Transmission'

In this supplemental material, we further analyze the dependence and robustness of our presented results on several assumptions and design parameters considered in the main text. First of all, as discussed in the main manuscript, we have considered increased losses in the gold plasmonic taper, assuming a collision frequency $\gamma_{tap} = 10\gamma$, where $\gamma$ is the collision frequency of bulk gold [19]. This was motivated by two fundamental issues: the effective collision frequency of a metal is modified (a) when the dimensions of a plasmonic waveguide become comparable to the electron mean free-path and (b) when the temperature of operation increases, leading to an effectively larger absorption coefficient. As we show in the following, this assumption does not qualitatively affect the overall concepts presented in the paper, other than the fact that a reduced absorption leads to a longer tapering length required to absorb the impinging energy.

In Fig. S1, we consider a plasmonic grating analogous to Fig. 1(a) of the main paper, but in which we have removed the entrance layer ($l = 0$) and assumed a collision frequency $\gamma_{tap} = \gamma = 17.65\,THz$ [30]. Its absorption features for $w = 6\,nm$, $d = 24\,nm$, $\varepsilon_S = 1$ and $l_{tap} = 1.76\,\mu m$, calculated for simplicity using our analytical TL model, validated with full-wave simulations for several examples, are shown in Fig. S2(a). The results are consistent with the ones obtained in the paper, after slightly increasing the length of the taper.

In order to decrease the required length $l_{tap}$ of the plasmonic taper we can use a more lossy plasmonic material, such as platinum (Pt), with experimental values of permittivity extracted from [1]. The absorption for an optimized design with $w = 24\,nm$, $d = 72\,nm$, $\varepsilon_S = 1$ and $l_{tap} = 980\,nm$ is shown in Fig. S2(b), providing similarly good performance.

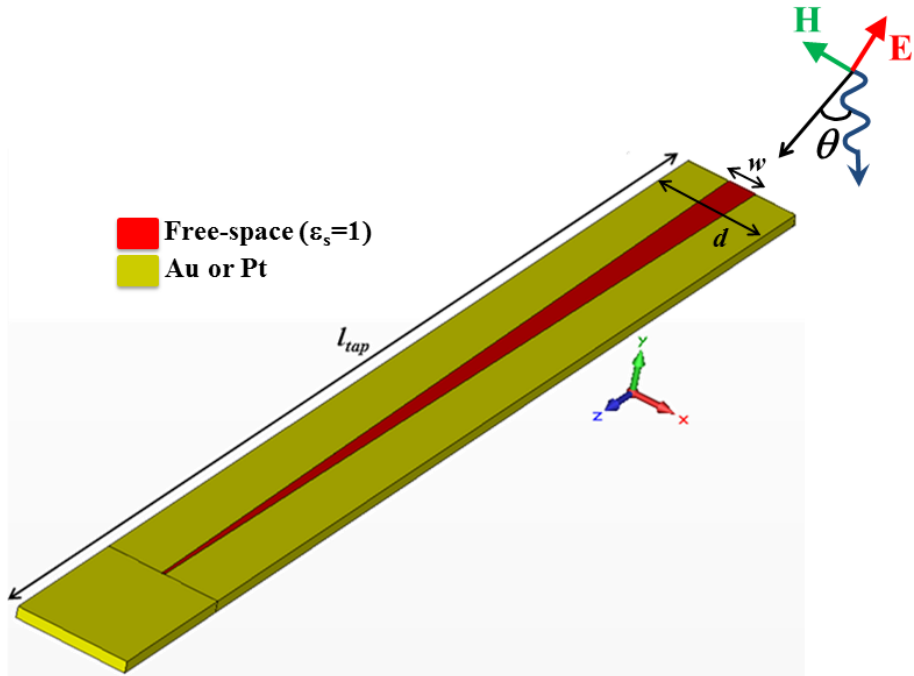

Figure S1 – Geometry of the 1D plasmonic taper. The device is illuminated by an oblique incidence TM polarized wave.

Finally, it may be desirable for some applications to extend the proposed plasmonic directional thermal emission to higher frequencies, such as near-infrared and optical. In order to achieve these selective emission properties, we need to increase the temperature to higher values than $T = 1000\,K$, for which gold will melt. Different metals needs to be considered for these cases with higher melting points, such as tungsten [1].

A directional thermal emitter based on tungsten is shown in Fig. S3(a) with dimensions $w = 6\,nm$, $d = 192\,nm$, $\varepsilon_S = 1$, $l = 200\,nm$ and $l_{tap} = 980\,nm$. The tungsten grating follows a relative Drude permittivity dispersion with parameters $\varepsilon_W = 1 - f_p^2 / [f(f+i\gamma)]$, $f_p = 1448$ THz, $\gamma = 13$ THz [1]. The taper has length $l_{tap}$ and is also made of tungsten, which follows a similar Drude model but with increased collision frequency $\gamma_{tap} = 130$ THz to take into account the temperature effects and the reduced slit width, similar to the previous discussion.

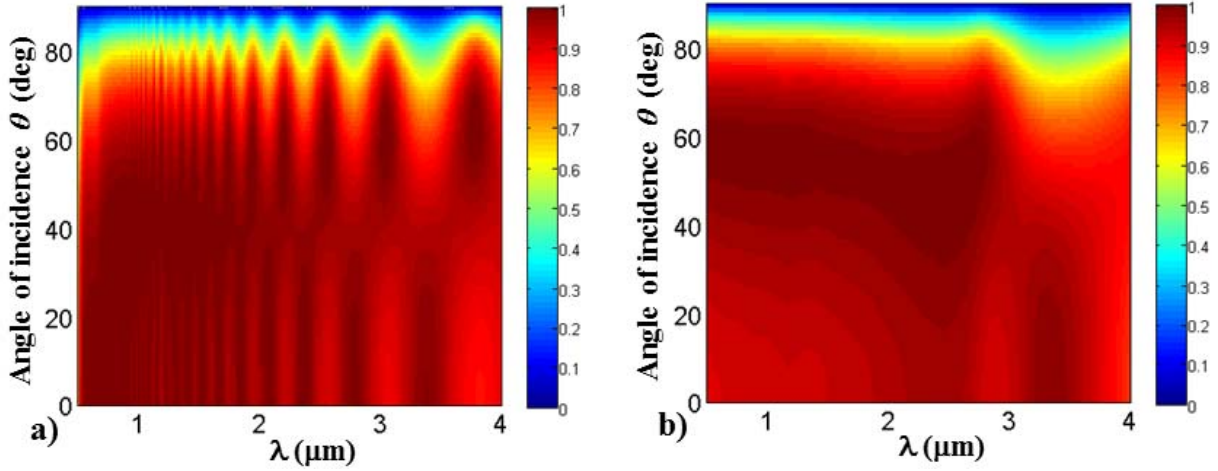

Figure S2 – a) Angular absorption spectra for the structure shown in Fig. S1 made of gold with period $d = 24\,nm$, slit width $w = 6\,nm$ and length $l_{tap} = 1.76\,\mu m$. b) Angular absorption spectra for the structure shown in Fig. S1 made of platinum with period $d = 72\,nm$, slit width $w = 24\,nm$ and length $l_{tap} = 980\,nm$. Both structures' slits are loaded with free-space permittivity $\varepsilon_S = 1$. Broadband omnidirectional absorption is obtained at optical and IR frequencies for both cases.

The selective thermal emission distribution at temperature $T = 1500\,K$ is shown in Figs. S3(b), centered at near-infrared frequencies. This may be further extended to optical frequencies in case we further increase the temperature, since the melting point of tungsten is $T = 3695\,K$. Broadband thermal emission with essentially the same bandwidth of an ideal blackbody (see inset in the bottom of Fig. S3(b)), is clearly obtained, confined to a narrow angular range with large spatial coherence. This directive near-infrared emission is confined around the plasmonic Brewster angle, as discussed in the manuscript.

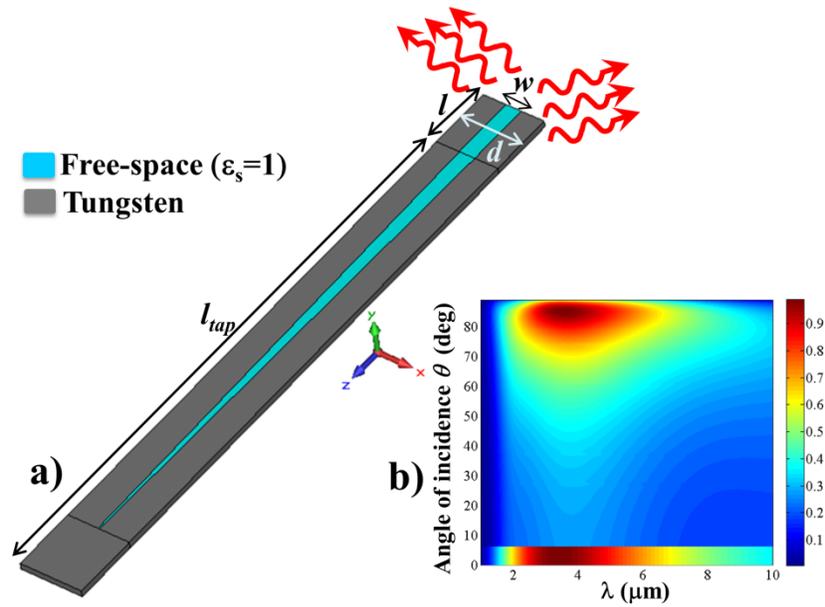

Figure S3 – a) Geometry of the 1D selective thermal emitter based on tungsten. b) Emission of the directional plasmonic emitter at $T = 1500\,K$ normalized to the maximum of the blackbody radiation at the same temperature, shown in the bottom part of (b).